\documentclass[12pt]{article}
\usepackage{amssymb}
\usepackage{epsfig}
\begin{document}
%
\setlength{\baselineskip}{0.65cm}
\setlength{\parskip}{0.35cm}
\renewcommand{\thesection}{\Roman{section}}
%
\begin{titlepage}
%
\vspace*{1.1cm}
\begin{center}
\LARGE

{\bf {Multiple parton interactions and  forward  \\[4mm]double pion
production in $pp$ and $dA$ scattering}}\\[2mm]
\vspace*{2.5cm}
\large 
{M.\ Strikman$^a$ and W.\ Vogelsang$^b$}

\vspace*{0.5cm}
\normalsize
\vspace*{0.5cm}
\normalsize
{\em $^a$Department of Physics, Pennsylvania State University,
University Park, PA, U.S.A.}\\

\vspace*{0.5cm}
{\em $^b$Institute for Theoretical Physics,
                Universit\"{a}t T\"{u}bingen,
                Auf der Morgenstelle 14,\\
                D-72076 T\"{u}bingen, Germany}\\

\end{center}

\vspace*{2.0cm}
\begin{abstract}
\noindent

We estimate the contributions by double-parton interactions 
to the cross sections for $pp\to \pi^0\pi^0X$ and $dA\to \pi^0\pi^0X$ 
at RHIC. We find that such contributions become important at large
forward rapidities of the produced pions. This is in particular the
case for $dA$ scattering, where they strongly enhance the azimuthal-angular 
independent ``pedestal'' component of the cross section, providing 
a natural explanation of this feature of the RHIC $dA$ data. 
We argue that the discussed processes open a window to studies of double
quark distributions in nucleons.
We also briefly address the roles of shadowing and energy loss
in $dA$ scattering, which we show to affect the double-inclusive 
pion cross section much more strongly than the single-inclusive one. 
We discuss the implications of our results for the interpretation of
pion azimuthal correlations.

\end{abstract}
\end{titlepage}
\newpage
%
\section{Introduction}

Cross sections for the production of identified hadrons at 
large transverse momentum play a crucial role at RHIC,
for both the spin and the heavy-ion program. In the latter they
serve as important probes of phenomena such as shadowing, gluon
saturation, or parton energy loss~\cite{rhicpapers}. For 
single-inclusive hadron production in $pp$ scattering, $pp\to h X$,
it was found that next-to-leading order (NLO) perturbative QCD~\cite{Werner} 
provides a very good description of the RHIC data over wide ranges
of transverse momentum, rapidity, and beam energy~\cite{d'Enterria:2006su}.
Striking suppression effects with respect to the $pp$ baseline have 
been observed on the other hand for scattering involving nuclei, 
among them in $dA$ scattering at forward 
rapidities~\cite{rhicfwd1,rhicfwd2,rhicfwd3}. 
These experimental studies at RHIC were also extended to the
production of two forward pions, in both $pp$ and $dA$ 
scattering~\cite{rhicfwd2,starqm09,phenixqm09}.
Of particular interest here are correlations between the pions in the
difference of their azimuthal angles, $\Delta
\varphi$. As expected, strong peaks in the 
distributions at $\Delta\varphi=0,\pi$ were observed in $pp$ scattering.
These are also present in peripheral $dA$ collisions. However, in
central $dA$ collisions, the ``backward'' peak at $\Delta\varphi \sim \pi$
is strongly depleted. It has been suggested that this depletion 
is due to gluon saturation effects in the Color Glass Condensate of 
the gold nucleus~\cite{Tuchin:2009nf,Albacete:2010pg}. 
At the same time, all distributions show a very 
significant $\Delta\varphi$-independent ``pedestal'' that
is much higher in the $dA$ case than in $pp$, a feature that has
received somewhat less attention. 

In this paper we will demonstrate that double-parton interactions for
which two leading partons of the ``projectile'' proton (or deuteron) 
interact with the ``target'' naturally make large contribution in 
the forward kinematics studied at RHIC, often dominating over
the leading-power contribution. They are particularly important
in $dA$ scattering. We will find that they could well be responsible 
for the pedestals in the $\Delta \varphi$ correlations and impact 
the interpretation of the observed correlations in both the pedestal
and the backward peak regions. 

Apart from their relevance for forward scattering at RHIC, 
double-parton interactions are also of wider interest in QCD 
as they provide a novel window on strong-interaction dynamics, 
including correlations of leading partons inside nucleons
or nuclei~\cite{Perugia}. 
As a result, they have received an ever growing attention over
the past few years~\cite{Strikman:2001gz,BDFS,anna,diehl}. In addition, 
understanding of 
double- and multi-parton interactions is also important for a
proper modeling of the structure of the final state for central 
$pp$ collisions at the LHC~\cite{Perugia,anna,lhc} and hence for the search for
new particles. Current experimental studies of multi-parton interactions 
involve selection of events with two back-to-back pairs 
of jets (or, a jet and a photon); see e.g.~\cite{Abe:1997xk,Abazov:2009gc}.
The fact that RHIC may provide a unique way to learn about 
multi-parton interactions, without having to use the more 
traditional double-scattering observables, is remarkable. 

Our paper is organized as follows. In Sec.~\ref{sect2} 
we discuss double-inclusive pion production in $pp$ scattering. 
We first demonstrate that in the leading-twist (LT) approximation the
cross section at forward rapidities involves incoming quarks with
very high momentum fraction. As a result, we find that double-parton 
processes in which two quarks each with relatively moderate $x\sim 0.3 
- 0.4$ scatter independently, become competitive over a fairly
wide kinematic range at RHIC. 
In Sec.~\ref{sect3} we study the double-parton mechanism for $dA$ 
collisions, which we find in the impulse approximation to be 
significantly enhanced as compared to the LT mechanism. We then 
discuss the impact of double-scattering contributions on the pedestal 
and the peak of the $\Delta\varphi$ distribution, along with generic 
features of gluon shadowing and parton energy loss, and argue that 
the suggested mechanisms allow to describe the bulk features of the data.
Finally, we summarize our results in Sec.~\ref{sect4}.

\section{Two-pion production in $pp$ scattering \label{sect2}}

In this section we explore the main features of $pp\to \pi^0 \pi^0 X$
through the LT mechanism based on a single hard scattering,
and through double-parton  interactions. We choose the case of $pp$ collisions,
both because  of its potential for studying new aspects of high energy 
QCD and nucleon correlation structure, but also because  it sets the 
baseline for our later discussion of nuclear scattering. In the following, 
$p_{T,1}$, $\eta_1$ are the transverse momentum and 
pseudo-rapidity of the ``trigger'' pion. The corresponding variables
of the second ``associated'' pion are denoted by $p_{T,2}$, $\eta_2$.
We consider collisions at center-of-mass energy $\sqrt{S}=200$~GeV
and $500$~GeV. 

\subsection{Leading-twist mechanism \label{sect21}}
We start with the LT mechanism for which two partons
collide in a single hard scattering. The generic expression for
the LT $pp\to \pi^0 \pi^0 X$
cross section is given in factorized form by
\begin{equation}
\frac{d^4\sigma_{\mathrm{LT}}}{dp_{T,1}d\eta_1dp_{T,2}d\eta_2} 
=\sum_{abcd}
\int dx_a dx_b dz_c dz_d \, f_a^p(x_a)f_b^p(x_b)\,
\frac{d^4\hat{\sigma}_{ab\to cdX}}{dp_{T,1}d\eta_1dp_{T,2}d\eta_2}\,
D_c^{\pi^0}(z_c)D_d^{\pi^0}(z_d) ,
\label{eq1}
\end{equation}
where the sum runs over all partonic channels, with $f_a^p$,
$f_b^p$ denoting the usual parton distribution functions of the
proton, $D_c^{\pi^0}$, $D_d^{\pi^0}$ the pion fragmentation functions
for partons $c$, $d$, and $\hat{\sigma}_{ab\to cdX}$ the
corresponding partonic hard-scattering cross sections. The latter
may be computed in QCD perturbation theory, starting at lowest 
order (LO) from $2\to 2$ scattering $ab\to cd$. Even though 
the NLO corrections are available 
in the literature~\cite{owens}, we will restrict ourselves for
this study to LO computations. We shall comment on this point 
below. We have for simplicity 
suppressed in Eq.~(\ref{eq1}) the dependence of the various functions
on the factorization/renormalization scale $\mu$. Throughout our
studies we choose the CTEQ6L parton distribution functions \cite{cteq6}
and the LO de Florian-Sassot-Stratmann (DSS) set of fragmentation 
functions \cite{DSS}. 

\begin{figure}[t]
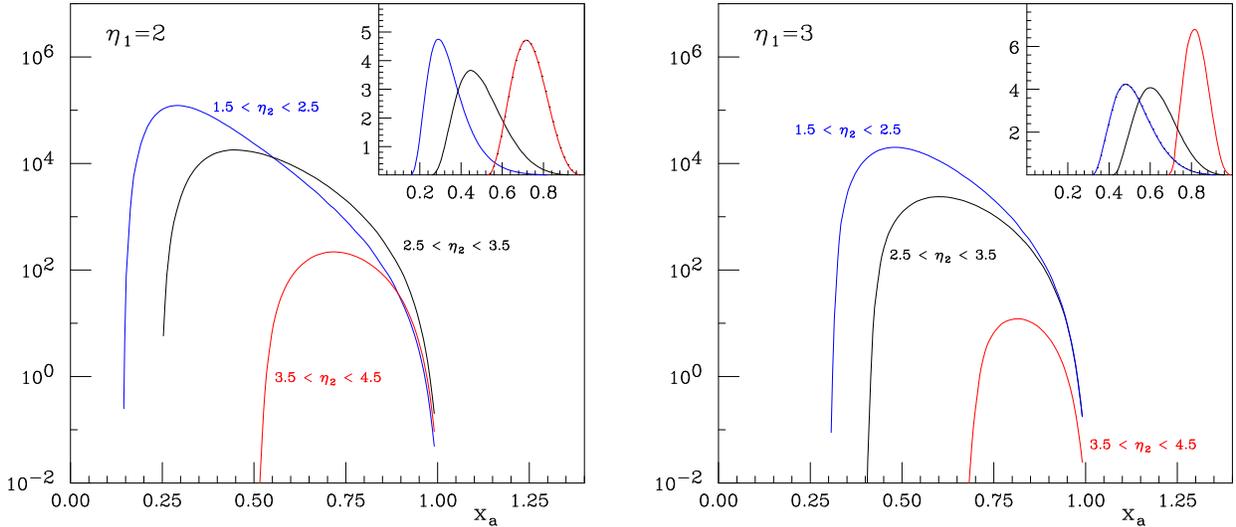
  
\psfig{file=x1histo1_lin.epsi, width=0.41\textwidth, angle=90}\hspace*{8mm}
\psfig{file=x1histo2_lin.epsi, width=0.41\textwidth, angle=90}
\caption{{\it Distributions of the leading-power LO cross section
for $pp\to \pi^0\pi^0X$  (see Eq.~(\ref{eq1})) at $\sqrt{S}=200$~GeV 
in momentum fraction $x_a$, for $\eta_1=2$ (left) and $\eta_1=3$ (right).
We have chosen $p_{T,1}=2.5$~GeV and integrated over 
$1.5~\mathrm{GeV}\leq p_{T,2}\leq p_{T,1}$ and various bins in $\eta_2$. 
Units are arbitrary. The inserts show the corresponding normalized
distributions on a linear scale.}}
    \label{xdistrfig}
 \end{figure}

Since $\eta_1+\eta_2=\log(x_a/x_b)$, 
production of two pions at relatively forward rapidities must
arise from ``imbalanced'' collisions where a large-$x$ parton 
from one proton hits a small-$x$ parton from the other~\cite{GSV}. 
These will typically be collisions of a valence quark and a gluon. 
Figure~\ref{xdistrfig} shows the distributions of the integrand
in Eq.~(\ref{eq1}) in $x_a$ for various values of $\eta_1$ and bins
in $\eta_2$. Here we have chosen $p_{T,1}=2.5$~GeV and $1.5~\mathrm{GeV}\leq 
p_{T,2}\leq p_{T,1}$ and the scale $\mu^2=(p_{T,1}^2 + p_{T,2}^2)/2$.
The inserts in the figure show the distributions on a linear scale, 
normalized in such a way that their integral is unity in each case.
One observes overall that with increasing $\eta_1$ or $\eta_2$ the 
distributions are shifted to higher values of $x_a$. In particular,
we see that the average value of $x_a=x_{\mathrm{quark}}$ 
for typical kinematics of the RHIC forward measurements is very high.

\subsection{Double-scattering mechanism}
The results shown in Fig.~\ref{xdistrfig} suggest that ``double-scattering'' 
contributions, with two separate hard interactions in a single $pp$ collision,
could become relevant at forward rapidities. Here the idea is that 
each of the two hard interactions produces a pion. The double-scattering 
contributions are power suppressed or ``higher-twist''. Like for 
the leading-twist contribution in Eq.~(\ref{eq1}), each of the two 
interactions will proceed primarily by a high-$x$ (valence) quark scattering 
off a small-$x$ gluon. However, compared to the leading-twist part,
the momentum fractions of the quarks participating in the double-scattering 
contributions will on average be much smaller, because for the latter the 
kinematics of the two recoiling ``unobserved'' partons 
is unconstrained. This makes the double-scattering 
contributions potentially dominant at forward rapidities.

A proper treatment of the double-parton mechanism would involve 
use of ``two-parton generalized parton distributions'' (2pGPDs) 
in the proton; see \cite{BDFS} for summary and references. This would allow to 
calculate the dimensional factor between single- and double-inclusive 
scattering, which characterizes the transverse spread of the double-parton 
distributions in the colliding
protons. At present, knowledge about 2pGPDs
is overall not sufficient to fully apply this formalism to the
case of hadron pair production in hadronic collisions. We therefore
resort to simple physically motivated estimates of the double-scattering
contribution. Here we are guided by the observation that in the case 
when the partons in each of the scattering protons are completely 
uncorrelated, we will have
\begin{equation}
\frac{d^4\sigma_{\mathrm{double,uncorr.}}}{dp_{T,1}d\eta_1dp_{T,2}d\eta_2} =
{1\over \pi R^2_{\mathrm{int}}}
\frac{d^2\tilde{\sigma}_{\mathrm{LT}}}{dp_{T,1}d\eta_1} 
\frac{d^2\tilde{\sigma}_{\mathrm{LT}}}{dp_{T,2}d\eta_2}
\label{eq2}
\end{equation}
for the double-scattering contribution to $pp\to\pi^0\pi^0X$, where
$\tilde{\sigma}_{\mathrm{LT}}$ denotes the leading-twist {\it single-inclusive}
cross section for $pp\to\pi^0X$, given by
\begin{equation}
\frac{d^2\tilde{\sigma}_{\mathrm{LT}}}{dp_Td\eta} =\sum_{abc}
\int dx_a dx_b dz_c  \, f_a^p(x_a)f_b^p(x_b)\,
\frac{d^2\hat{\sigma}_{ab\to cX}}{dp_Td\eta}\,
D_c^{\pi^0}(z_c) ,
\label{eq3}
\end{equation}
with single-inclusive partonic cross sections 
$\hat{\sigma}_{ab\to cX}$. 
Furthermore, in Eq.~(\ref{eq2}) $  \pi R^2_{\mathrm{int}}$ 
is an ``effective'' transverse area covered by the two correlated partons
(it was denoted as $\sigma_{\mathrm{eff}}$ in a number of experimental papers 
and some of the theoretical papers, although it has little to do with an
interaction cross section).
In the approximation of partons uncorrelated in the transverse plane it 
can be expressed through a convolution of usual 
generalized parton distributions in the 
hadrons \cite{BDFS,Frankfurt:2003td}.
Hence, if we assume for simplicity that the partons' transverse spread 
does not depend on their momentum fractions $x$, we can write
\begin{eqnarray}
\frac{d^4\sigma_{\mathrm{double}}}
{dp_{T,1}d\eta_1dp_{T,2}d\eta_2} &=&{1\over  \pi 
R^2_{\mathrm{int}}}
\sum_{abc\,a'b'c'}
\int dx_a dx_b dz_c dx_{a'} dx_{b'} dz_{c'} \, 
f_{aa'}^p(x_a, x_{a'})f_b^p(x_b)f_{b'}^p(x_{b'})\nonumber 
 \\[2mm]
&\times &
\frac{d^2 \hat{\sigma}^{ab\to cX}}{dp_{T,1} d\eta_1}\,
\frac{d^2 \hat{\sigma}^{a'b'\to c'X'}}{dp_{T,2} d\eta_2}\,
D_c^{\pi^0}(z_c)\,D_{c'}^{\pi^0}(z_{c'})\,.
\label{eq4}
\end{eqnarray}
Here  $f_{aa'}^p(x_a, x_{a'})$ is a ``double-parton'' distribution for
partons $a,a'$ in the same proton, which we will model in the following. 
If the partons are not correlated, it is equal to the product of two
ordinary parton distributions, $f_{aa'}^p(x_a, x_{a'})=f_a^p(x_a)
f_{a'}^p(x_{a'})$, and Eq.~(\ref{eq4}) reverts to (\ref{eq2}). 
As shown in Eq.~(\ref{eq4}), we neglect for simplicity correlations 
in the ``target'', i.e., in the proton probed at small-$x$.

The double-parton distribution has to obey the kinematic constraint
$x_a + x_{a'}\le 1$. We implement this condition for all partonic
channels. Beyond that, we only consider double-parton correlations 
for valence quarks, that is, for the case $a,a'\equiv q,q'$, 
with $q=u,d$. For these we make the ansatz
\begin{equation}
f_{qq'}^p(x_q,x_{q'})= \frac{1}{2} \left[ f_q^p(x_q)\times \phi
\left(\frac{x_{q'}}{1-x_q}\right)+\left( q\leftrightarrow q'\right)\right].
\label{corr}
\end{equation}
The picture we have in mind here is that the first interaction involves
a 
 valence quark with its distribution $f_q(x_q)$. The distribution
of a second valence quark that participates in the second hard
scattering is then expected to be modified relative to the 
usual parton distribution. For instance, if the first hard scattering
involves an up quark, then fewer up quarks will be available for
the second interaction. We assume this effect to be described by 
a single function $\phi$, given by 
\begin{equation}
\phi(\xi)=\frac{c}{\sqrt{\xi}}(1-\xi)^n,
\label{corr1}
\end{equation}
with $c=3/4$ and $n=1$. The latter value follows from counting rule
arguments; scaling violations would be expected to increase it somewhat. 
The normalization factor $c$ in Eq.~(\ref{corr1}) is determined 
from the baryon number sum rule $\int_0^1d\xi \phi(\xi)=1$. 
Since the expression for the cross section is symmetric in 
$x_q,x_{q'}$ we perform symmetrization of Eq.~(\ref{corr}). For
all partonic combinations not involving valence quarks, we also 
use Eq.~(\ref{corr}), but with $\phi$ replaced by the 
usual parton distribution function $f_{a'}^p(x_{a'}/(1-x_a))$. Here
the modified argument guarantees that the kinematic constraint
$x_a + x_{a'}\le 1$ is respected. Our procedure should be compared 
to the model of \cite{Sjostrand:2004pf} where it was assumed that 
also for valence quarks the function $\phi(\xi) $ is given by the 
usual distribution $f_q^p(\xi)$, which has $n\sim 3$,
and no symmetrization was performed. While our ansatz
arguably is physically better motivated, we do not find much numerical
difference between the two models, except very close to the phase
space boundary at high rapidities and/or transverse momenta. 
Rather than the precise choice of $\phi$, it is the kinematic 
constraint $x_a + x_{a'}\le 1$ that matters most in our numerical studies,
reducing the cross section.

For our calculations, we choose  $\pi R^2_{\mathrm{int}}
= 15$~mb~\cite{Abe:1997xk,Abazov:2009gc}  
in Eq.~(\ref{eq2}). This experimental value is smaller than 
$\pi R_{\mathrm{int}}^2= 34$~mb obtained in the mean field approximation 
for the 2pGPDs when partons are not correlated in the 
transverse plane~\cite{Frankfurt:2003td}. 
The value $\sim$ 34 mb  is  an upper limit on 
 $\pi R_{\mathrm{int}}^2$ when only momentum fractions larger than 
$\sim 10^{-2}$ are relevant, provided there is no 
repulsion between the partons. A smaller experimental value 
of $\pi R_{\mathrm{int}}^2$
indicates the presence of transverse correlations among partons. 
In principle these correlations can depend on the flavors and momentum
fractions of the partons involved in the double-scattering 
interaction. Qualitatively, we expect that picking two leading 
quarks would select configurations with reduced transverse 
separation  between the quarks, 
making it more natural to use the experimental value for 
$\pi R^2_{\mathrm{int}}$   than the uncorrelated estimate. 
It is of interest here that in the limit when the typical 
transverse separation between the quarks is much smaller than for the 
small-$x$ gluons, one can derive based on~\cite{BDFS} 
\begin{equation}
\pi R^2_{\mathrm{int}}= \left[\int {d^2 \Delta \over (2\pi)^2}  
F^2_{2g}(\Delta)\right]^{-1}={12\pi \over m_g^2} \approx \mbox{14 mb}.
\end{equation}
Here $F_{2g}(\Delta) \approx 1/(\Delta^2/m_g^2+1)^2$ with
$m_g^2(x\sim 0.01)=1.1 \, \mbox{GeV}^2$ is the two-gluon form factor 
of the nucleon. A larger value of $\pi R^2_{\mathrm{int}}$ would evidently 
reduce the size of our estimates for the double-interaction contribution.

\begin{figure}[p]
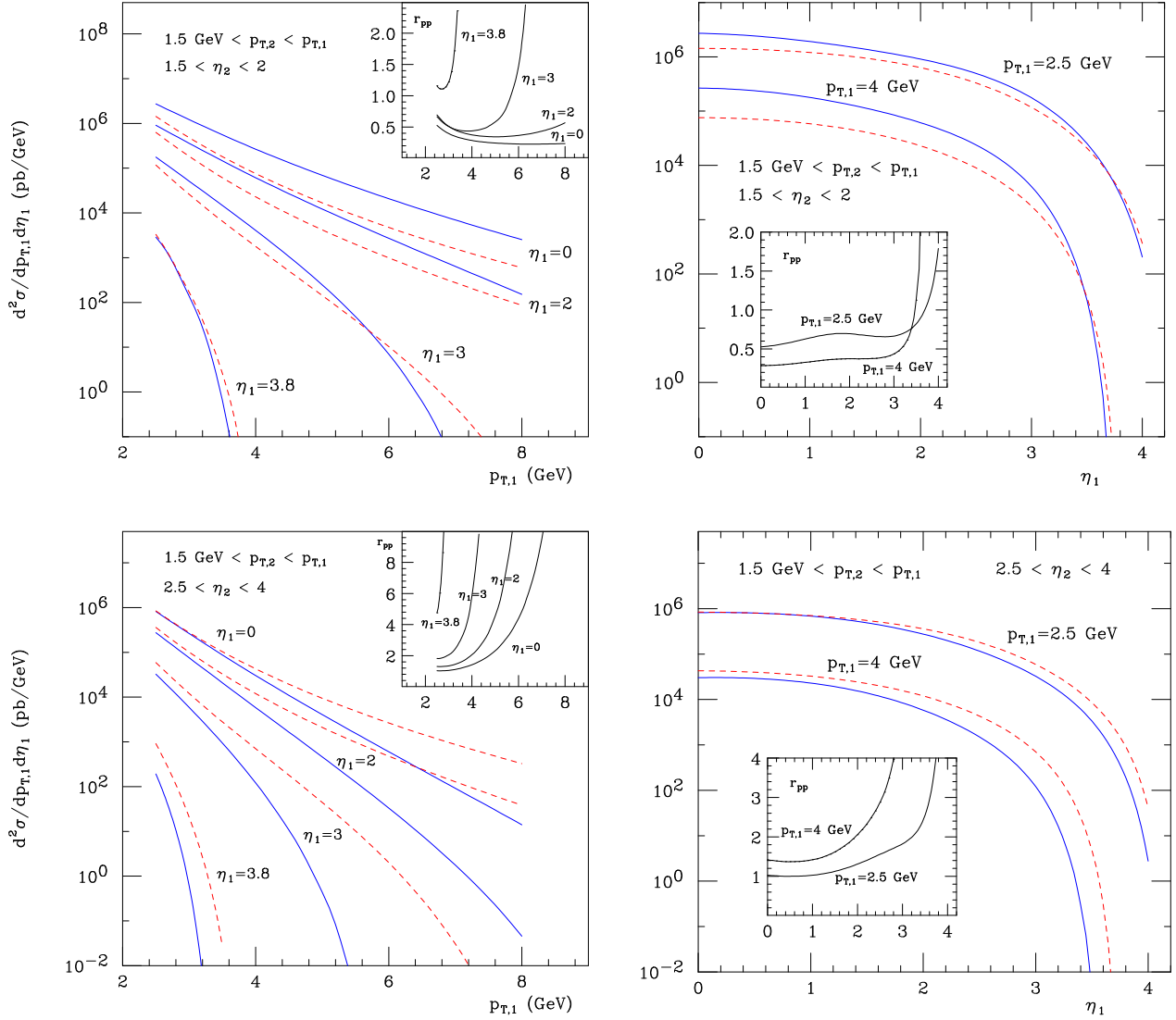
  
\psfig{file=sigma_2bins.epsi, width=0.41\textwidth, angle=90}\hspace*{8mm}
\psfig{file=sigma_2bins_eta.epsi, width=0.41\textwidth, angle=90}\\[6mm]
\psfig{file=sigma_2bins1.epsi, width=0.41\textwidth, angle=90}\hspace*{6mm}
\psfig{file=sigma_2bins1_eta.epsi, width=0.41\textwidth, angle=90}
   \caption{{\it Comparison of the leading-twist cross section for
$pp\to \pi^0\pi^0X$ (solid, see Eq.~(\ref{eq1})) and the 
double-interaction contribution estimated from Eq.~(\ref{eq4})
(dashed), as functions of $p_{T,1}$ (left) and $\eta_1$ (right).
The plots in the upper row are for $1.5< \eta_2 < 2$, the ones
in the lower row for $2.5< \eta_2 < 4$. For all plots we have 
chosen $1.5~\mathrm{GeV}<p_{T,2}<p_{T,1}$. The inserts in each plot
show the ratio $r_{pp}$ of the double-interaction contribution to 
the leading-twist one, see Eq.~(\ref{eq7}).}}
    \label{MIfig}
 \end{figure}

Figure~\ref{MIfig} shows our results for the leading-twist 
cross section $d\sigma_{\mathrm{LT}}$ for 
$pp\to \pi^0\pi^0X$ in Eq.~(\ref{eq1}) and
for the double-interaction contribution 
$d\sigma_{\mathrm{double}}$ according to Eq.~(\ref{eq4}), 
as functions of the trigger pion's transverse momentum and
rapidity. For the associated pion we have integrated the cross sections 
over $1.5~\mathrm{GeV}\leq  p_{T,2}\leq p_{T,1}$ and $1.5\leq \eta_2\leq 2$
(upper row) or $2.5\leq \eta_2\leq 4$ (lower row).
$d\sigma_{\mathrm{LT}}$ has been calculated as before; for 
$d\sigma_{\mathrm{double}}$ in Eq.~(\ref{eq4}) we have chosen the same parton
distributions and fragmentation functions, and the scales $\mu=p_{T,i}$.
All calculations are done at LO. 
As one can see from Fig.~\ref{MIfig}, the estimated double-scattering
contribution shows the typical features of a higher-twist 
(power-suppressed) contribution. It tends to increase relative to
the leading-twist cross section towards lower transverse momenta.
Near mid-rapidity and for moderately high $p_{T,1}$, 
double-scattering is essentially negligible.
On the other hand, it also increases towards the kinematic boundaries
at high rapidities and transverse momenta. Therefore, it is likely
to play a significant role for much of the kinematic regime relevant in
the studies of two-pion correlations at forward rapidities at RHIC. Here
it would affect also the distributions in the difference $\Delta\varphi$ 
of the azimuthal angles of the two pions, where it should enhance both 
the backward peak at $\Delta\varphi\sim\pi$ and the ``pedestal'' at 
$\Delta\varphi<\pi$. It is worth emphasizing already at this point that 
the lowest-order LT part only contributes at $\Delta\varphi=\pi$,
whereas the double-scattering piece will uniformly
contribute at all $\Delta\varphi$. Hence, as the LT cross section 
receives contributions at $\Delta\varphi<\pi$ only at higher orders in 
perturbation theory, the double-scattering is 
expected to dominate even more strongly away from the backward peak. 
For future reference, we also define the ratio 
of the double-scattering contribution to the leading-twist one:
\begin{equation}
r_{pp}\equiv \frac{d^4\sigma_{\mathrm{double}}}{d^4\sigma_{\mathrm{LT}}} .
\label{eq7}
\end{equation}
We show the results for $r_{pp}$ in the inserts in Fig.~\ref{MIfig}.
This number should be compared with the ratio of the areas under 
the pedestal and under the backward peak at $\Delta\varphi\sim \pi$, 
which is of the order of two.

\begin{figure}[t]
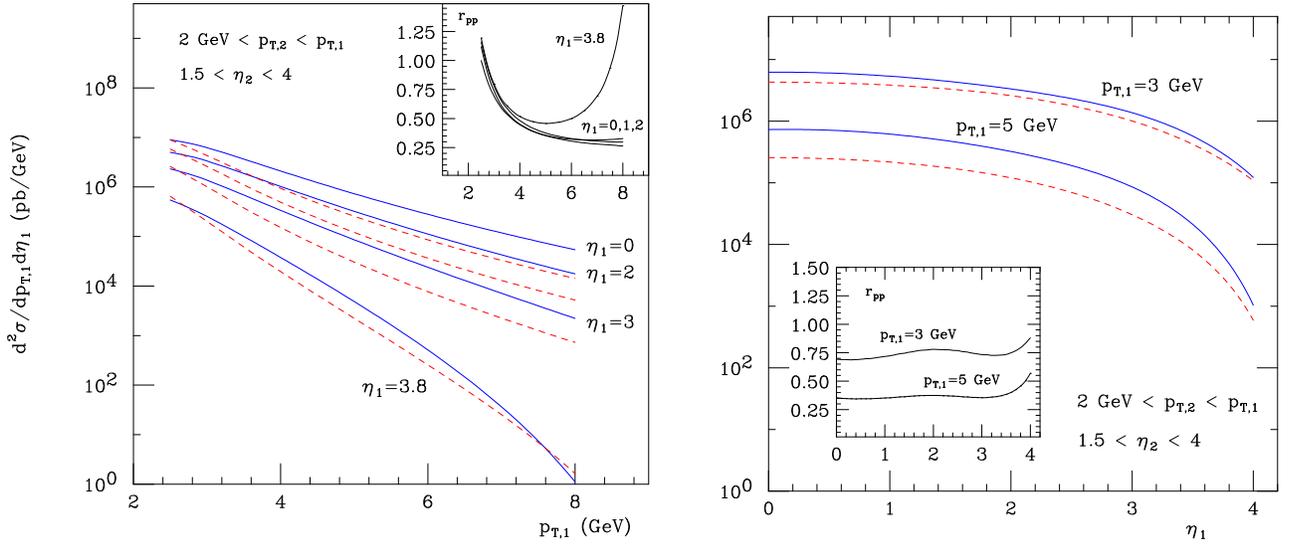
  
\psfig{file=sigma_500.epsi, width=0.42\textwidth, angle=90}\hspace*{8mm}
\psfig{file=sigma_eta_500.epsi, width=0.41\textwidth, angle=90}
   \caption{{\it As Fig.~\ref{MIfig}, but at $\sqrt{S}=500$~GeV
and for $2~\mathrm{GeV}<p_{T,2}<p_{T,1}$ and $1.5< \eta_2 < 4$.}}
    \label{MIfig500}
 \end{figure}

As there will be high luminosity runs at RHIC with polarized protons 
at $\sqrt{S}=500$~GeV, we have also performed calculations at this energy. 
Figure~\ref{MIfig500} shows the results as functions of $p_{T,1}$ and 
$\eta_1$, integrated over $2~\mathrm{GeV}<p_{T,2}<p_{T,1}$ and $1.5<\eta_2<4$. 
As expected the effects are overall smaller than at $\sqrt{S}=200$~GeV, but
remain significant at large rapidities. 

As we have hinted at earlier, there are still considerable uncertainties
in the computation of the double-scattering contribution. We remind
the reader that one would in principle need to set up a framework based on 
2pGPDs. Even within our ansatz in Eq.~(\ref{eq2}) there is some
uncertainty regarding the value for $\pi R_{\mathrm{int}}^2$ and the
model used for the double-parton correlation functions. 
Furthermore, at forward rapidities the fragmentation functions are
probed at rather high momentum fractions $z$, where they are not
known accurately. On top of
this, one needs to address the role of higher-order QCD corrections.
The double-inclusive and single-inclusive leading-twist cross sections
carry significant dependence on the renormalization/factorization
scales. 
While the NLO corrections are available for both the double-inclusive
leading-twist cross section~\cite{owens} and for the single-inclusive
one in Eq.~(\ref{eq3})~\cite{Werner}, it is not guaranteed that the
form of $d\sigma_{\mathrm{double}}$ in Eq.~(\ref{eq4}) carries over
to higher orders of perturbation theory, since particle radiation
will tend to correlate the two separate hard interactions
(this effect should be small, however, for configurations which 
dominate in the mean field uncorrelated approximation, since in 
this case the bulk of the parton cross sections originates from quark 
transverse separations much larger than $1/p_T$). That said,
each of the cross sections  $d\sigma_{\mathrm{LT}}$, 
$d\tilde{\sigma}_{\mathrm{LT}}$ in Eqs.~(\ref{eq1}),(\ref{eq2}) 
is known to receive positive NLO radiative corrections of 
$\gtrsim 50\%$ or so for RHIC kinematics, so that it appears likely 
that {\it QCD corrections will overall enhance the relevance of the 
double-scattering contribution}.

The uncertainties inherent in the present calculations 
somewhat limit the possibilities to achieve
a better determination of $\pi R_{{\mathrm{int}}}^2$ from RHIC measurements. 
Nonetheless, if our phenomenological predictions are correct, 
it might be possible to identify the double-scattering contributions
from detailed studies of the dependence of the two-pion cross section
on transverse momenta and rapidities. A further 
possible test of this picture would be to measure a third pion at 
``recoil kinematics'' $\eta_3\sim 0$. 
This could serve to further
enhance the double-scattering contribution over the leading-twist one,
since the latter can give rise to a third pion only at higher orders
in perturbation theory, whereas the double interactions naturally
give rise to a third (and even a fourth) recoiling ``jet''. 
Obviously the study of polarization effects in two-pion correlations 
would be of interest as well in the context of the double-scattering
mechanism.

\section{Two-pion production in $dA$ scattering \label{sect3}}

\subsection{Introductory remarks \label{3.1}}

An important finding at RHIC~\cite{rhicfwd1,rhicfwd2,rhicfwd3} 
is that the rate of forward 
pion production at relatively large transverse momenta, where perturbative 
QCD describes the corresponding $pp$ data, is suppressed in $dA$ scattering 
by a large factor as compared to the impulse approximation result. 
This suppression is expressed by the ``nuclear modification factor''
$R_{dA}$, which effectively compares the observed production rates for a 
given centrality trigger to the prediction based on the approximation that 
the parton density in nuclei at an impact parameter $b$ is equal to 
the additive sum of the parton densities of individual nucleons at 
this impact parameter. 
A  more formal way to formulate the latter assumption 
is to define the impact-parameter dependent parton distribution
of the nucleus, $f_a^A(x,Q^2, b)$, which coincides with the 
corresponding diagonal generalized parton distribution (GPD)
in impact parameter representation~\cite{Burkardt}.
In the discussed approximation, $f_a^A(x,Q^2, b)$ is given by 
 \begin{equation}
 f_a^A(x,Q^2, b) = f_a^N(x,Q^2) T_A(b),
 \end{equation}
where  $T_A(b)$ is the standard nuclear profile function which
is given by an integral of the density function over the
longitudinal direction:
\begin{equation}
T(b)=\int \rho_A\left(\sqrt{b^2+z^2}\right) dz. 
\label{tab}
\end{equation}
$T_A$ is normalized to $\int d^2b T(b)=A$. 
The experimental data show that the 
suppression becomes stronger with increase of rapidity $\eta$. It is 
found that $R_{dA}$ is typically of the order $1/3$ for forward kinematics.
Furthermore, the suppression becomes stronger 
with decrease of $b$ and is strongest for $b\sim 0$.

The analysis~\cite{GSV} has demonstrated that the dominant mechanism 
for single-inclusive pion production in the forward kinematics explored
at RHIC is scattering of a leading quark of one (``projectile'') nucleon 
off a gluon in the other (``target'') nucleon. The median value of 
momentum fraction $x_g$ of the gluon was found to be in the range 
$x_g \sim 0.01 - 0.03$, depending on the rapidity of the pion.
The nuclear gluon density for such values of $x_g$ is known to be 
close to the incoherent sum of the gluon fields of the individual 
nucleons since the coherence length in the interaction is rather 
modest for the distances involved. As a result, the leading-twist  
nuclear shadowing effects cannot explain the observed 
suppression~\cite{GSV}, and one needs a novel dynamical mechanism 
to explain the suppression of pion production in such collisions. 
 
An important additional piece of information comes from the study 
of correlations of the leading forward pion with an additional pion
produced at central rapidities~\cite{rhicfwd2,centrfwd}. In this case,
the dominant contribution comes from the scattering off gluons with 
$x_g\sim 0.01 - 0.02$. An extensive analysis performed in \cite{FS07} 
has demonstrated that the strengths of such forward-central 
correlations are similar in $dA$ and in $pp$ scattering
once one corrects for the contribution of soft interactions 
to the pion yield at $\eta \sim 0$,
and that in $dA$ the dominant source of leading pions is scattering 
at large impact parameters. This conclusion is supported by the 
observation of the STAR experiment~\cite{Rakness} that the 
associated multiplicity of 
soft hadrons in events with a forward pion is a factor of two smaller 
than in minimum-bias $dA$ events. This reduction factor is consistent 
with the estimate of~\cite{FS07}. Overall, the patterns
observed in forward inclusive-pion production and forward-central 
correlations are consistent with the picture of effective energy losses  
which we further discuss in Sec.~\ref{3.3}. Hence we will use it below 
for the numerical estimates of the deviation form the impulse approximation.
We note in passing that the above mentioned features of the forward 
pion production data represent a challenge for the $2\to 1 $ scattering 
mechanism \cite{Kharzeev,Dumitru:2005gt} 
that dominates in the color glass condensate model. In this mechanism  
forward pions are predominantly produced at central impact parameters 
without producing recoil pions  at central rapidities.
  
In Ref.~\cite{GSV} we suggested that to study the effects of small-$x$ 
gluon fields in the initial state one needs to study production of {\it two
leading forward} pions in nucleon$-$nucleus collisions. Recently such data were 
taken in $dAu$ collisions~\cite{rhicfwd2,starqm09,phenixqm09}. In the next 
subsection, we will analyze the role of the double-parton interactions 
in the kinematics explored at RHIC. The effects associated with 
suppression of the single-inclusive spectrum mentioned above will be 
discussed further below in Sec.~\ref{3.3}.
 
 \subsection{Double-parton versus single-parton interactions $-$ 
treatment in the impulse approximation  \label{3.2}}

As we saw in the previous section, large values of the rapidities 
of the two pions select high $x$ in the ``projectile'' hadron, and
double-scattering contributions may become very significant. 
As measurements in this kinematic domain have been carried out
in $dAu$ scattering at RHIC~\cite{rhicfwd2,starqm09,phenixqm09}, 
it is of much interest to see in how
far the double-interaction contributions are further enhanced in the 
reaction $dA\to \pi^0\pi^0X$. Compared to the $pp$ case, it is clear
that the presence of many nucleons in the scattering process will
offer more possibilities for multiple-parton interactions.

\begin{figure}[t]
   \centering
   \includegraphics[width=1.0\textwidth]{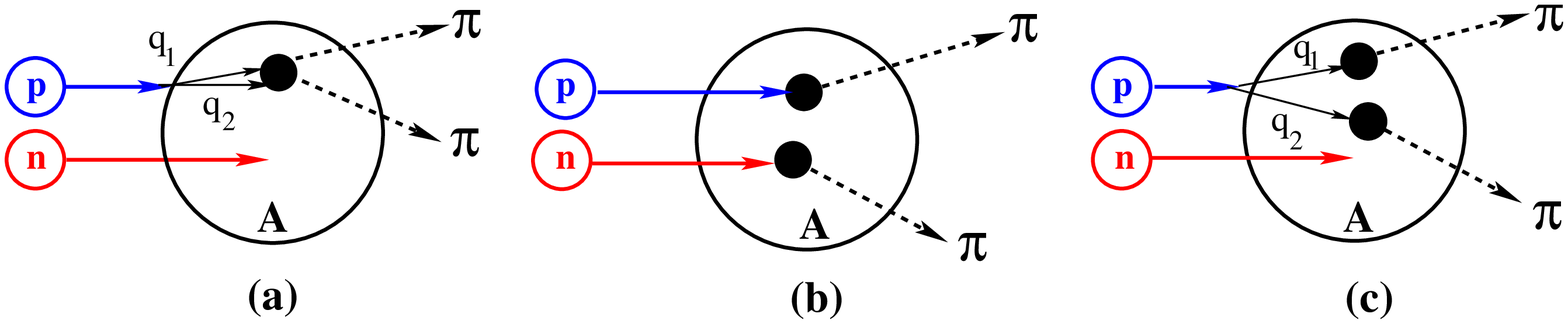} 
   \caption{{\it Contributions to two-pion production in $dA$ collisions
through the double-interaction mechanism.}}
    \label{abc}
 \end{figure}

We may distinguish three contributions to the double-parton
mechanism in $dA$ scattering, as shown in Fig.~\ref{abc}:
\begin{description} 
\item[(a)] Two (valence) quarks from one of the nucleons in the deuteron 
participate in the hard-scattering, striking the same nucleon in
the heavy nucleus (Fig.~\ref{abc}(a)).
\item[(b)] Independent scattering of the deuteron's proton and neutron
off separate nucleons in the heavy nucleus. Each of the two 
collisions produces one of the observed pions (Fig.~\ref{abc}(b)).
\item[(c)] Same as (a), but with the double interaction occurring off 
two different nucleons in the  heavy nucleus. Again each of the two collisions 
produces one of the observed pions (Fig.~\ref{abc}(c)). 
\end{description}
We now proceed to make estimates for these contributions. For our more 
illustrative purposes, we neglect effects of nuclear (anti-)shadowing 
for the heavy nucleus. Also, we treat the heavy nucleus as roughly
iso-scalar. For our estimates we need to take into account the 
distribution of nucleons in a heavy nucleus. Since the experiments 
are performed with a centrality trigger, it is useful to first 
write the double-inclusive cross section in a form where the 
integral over impact parameter $b$ is kept explicitly~\cite{Strikman:2001gz}. 
We write all expressions for $N$-nucleus scattering,
where $N=(p+n)/2$ denotes an iso-scalar combination of proton and
neutron. Since they are bound in a deuteron they 
propagate at similar impact parameters. 
We further assume that the impulse approximation is  
valid  for the interaction with the nucleus. For any contribution that
involves scattering off only one of the ``target'' nucleons, we 
then have the generic formula 
\begin{equation}
\frac{d^4\sigma^{NA}}{dp_{T,1}d\eta_1dp_{T,2}d\eta_2}= 
\int d^2b\; T(b) 
\frac{d^4\sigma^{NN}}{dp_{T,1}d\eta_1dp_{T,2}d\eta_2}
\label{impulsedouble1}
\end{equation}
for the two-pion cross section. Here, $T(b)$ is the nuclear 
thickness factor defined above in Eq.~(\ref{tab}).
Equation~(\ref{impulsedouble1}) holds for
contribution (a), but evidently also for the leading-twist piece. 
Hence, if we consider a fixed impact parameter and take their ratio,
the factor $T(b)$ will cancel:
\begin{equation}
r_{\mathrm{a}}(b) \equiv \frac{d^4\sigma^{NA}_{\mathrm{double,(a)}}}
{d^4\sigma^{NA}_{\mathrm{LT}}}=\frac{d^4\sigma^{NN}_{\mathrm{double}}}
{d^4\sigma^{NN}_{\mathrm{LT}}}=r_{NN}.
\label{rc}
\end{equation}
In the last step we have used our definition in Eq.~(\ref{eq7}) 
for the ratio of the double-interaction contribution to the 
leading-twist one, now adapted to the case of $NN$ collisions.
Apart from trivial (and small) isospin modifications 
related to the fact that there are contributions here from 
$pp$, $pn$, $np$, and $nn$ scattering, $r_{NN}$ is identical to 
$r_{pp}$ considered in the previous section. 

The situation is different, however, for contributions (b) and (c),
for which two ``target'' nucleons are involved in the scattering,
so that the square of $T(b)$ will appear in the 
expressions~\cite{Strikman:2001gz}. For contribution (b) we have
\begin{equation}
\frac{d^4\sigma^{NA}_{\mathrm{double,(b)}}}{dp_{T,1}
d\eta_1dp_{T,2}d\eta_2}= \int d^2b \;
 {A-1\over A}\;T^2(b) \times\frac{1}{2}\left[
\frac{d^2\tilde{\sigma}^{pN}_{\mathrm{LT}}}{dp_{T,1}d\eta_1} 
\frac{d^2\tilde{\sigma}^{nN}_{\mathrm{LT}}}{dp_{T,2}d\eta_2}+\left( 
p_{T,1},\eta_1 \leftrightarrow p_{T,2},\eta_2 \right)\right],
\label{impulsedouble}
\end{equation}
where $\tilde{\sigma}_{\mathrm{LT}}$ 
again denotes a LT single-inclusive cross
section as introduced in Eq.~\ref{eq3}. Here
we have neglected for simplicity any correlations between
quarks in the two ``projectile'' nucleons. As indicated,
we need to properly symmetrize (\ref{impulsedouble}), 
since either the $pN$ or 
the $nN$ interaction can produce a given pion. Taking again the
ratio to the leading-twist term at fixed impact parameter, 
one factor of $T(b)$ cancels, and we have for large $A$:
\begin{equation}
r_{\mathrm{b}}(b)=
\frac{d^4\sigma^{NA}_{\mathrm{double,(b)}}}
{d^4\sigma^{NA}_{\mathrm{LT}}}=
\frac{T(b) \left[ d^2\tilde{\sigma}^{pN}_{\mathrm{LT}} 
d^2\tilde{\sigma}^{nN}_{\mathrm{LT}}+\left( 
p_{T,1},\eta_1 \leftrightarrow p_{T,2},\eta_2 \right)\right]}
{2\,d^4\sigma^{NN}_{\mathrm{LT}}}.
\label{ra}
\end{equation}
Finally, for contribution (c) we define analogously
\begin{equation}
r_{\mathrm{c}}(b) \equiv \frac{d^4\sigma^{NA}_{\mathrm{double,(c)}}}
{d^4\sigma^{NA}_{\mathrm{LT}}}.
\end{equation}
Here the numerator is again proportional to $T^2(b)$, while the
denominator is linear in $T(b)$. With the help of Eq.~(\ref{eq4})
we find at fixed impact parameter:
\begin{equation}
r_{\mathrm{c}}(b) = T(b)\,\pi R_{\mathrm{int}}^2
\,
\frac{d^4\sigma^{NN}_{\mathrm{double}}}
{d^4\sigma^{NA}_{\mathrm{LT}}}=T(b)\,\pi R_{\mathrm{int}}^2\,r_{NN}\;,
\label{rb}
\end{equation}
again up to small isospin corrections. 

Before presenting more detailed numerical results for the ratios
in Eqs.~(\ref{rc}),(\ref{ra}),(\ref{rb}), we discuss their relative 
size. From (\ref{rc}),(\ref{rb}) we see immediately that to
very good approximation
\begin{equation}
\frac{r_a}{r_c}=\frac{1}{T(b)\,\pi R_{{\mathrm{int}}}^2}\;.
\end{equation}
As before we use $\pi R_{\mathrm{int}}^2= 15$~mb. For heavy nuclei
with $b\sim 0$ we have $T(0)\approx 2.2$~fm$^{-2}=1/(4.54~\mathrm{pb})$.
Therefore, we have $r_a/r_c\approx 0.3$. The ratio of $r_b$ and $r_c$ will 
be close to one at mid-rapidity where correlations and valence-gluon
scattering are not very important. Towards large rapidities, however,
$r_b$ must become much larger than $r_c$, since it is not
subject to the constraint $x_a+x_{a'}\leq 1$ because of the fact that
for (b) the proton and the neutron scatter independently. 

\begin{figure}[p]
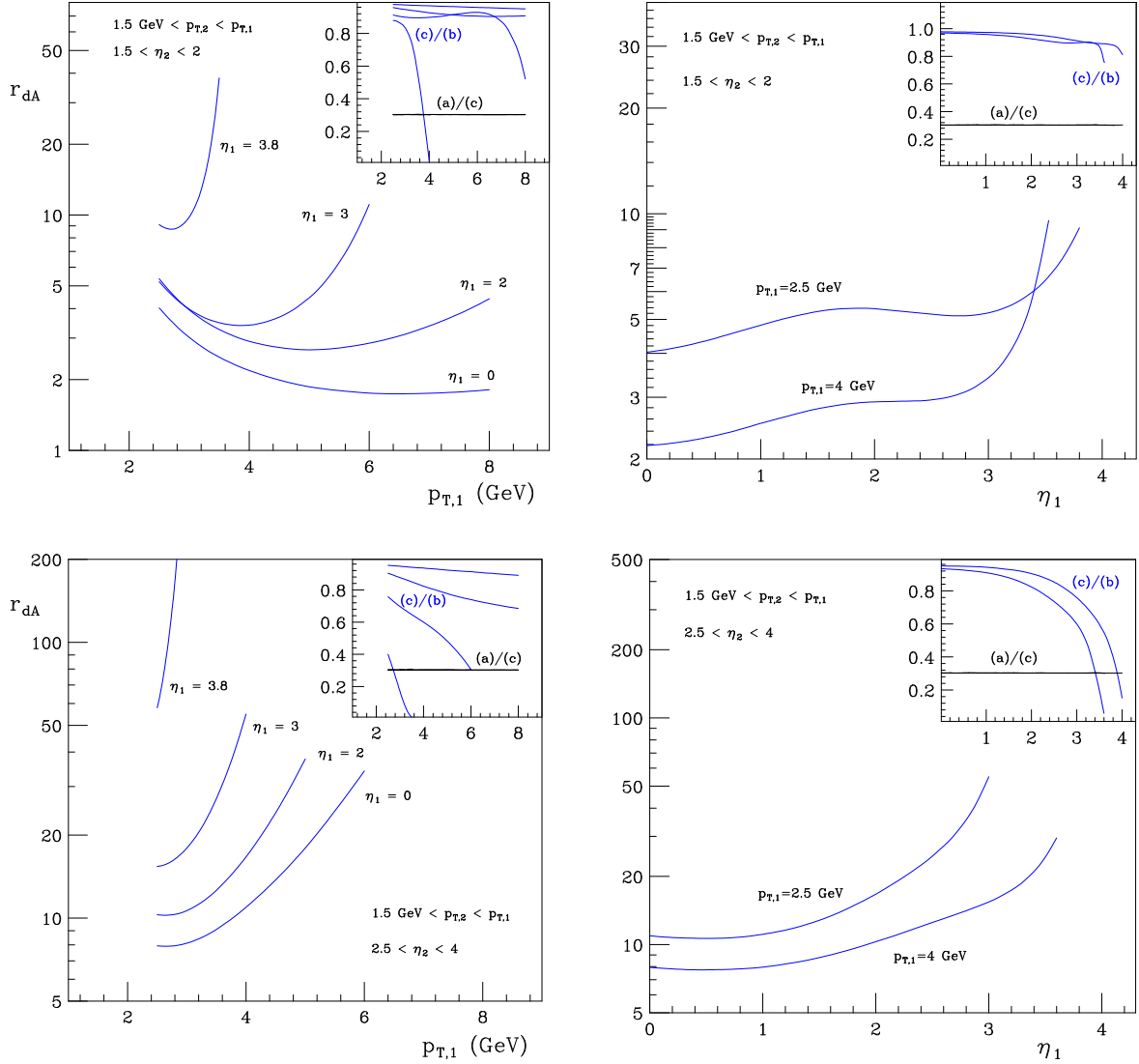
  
\psfig{file=rabc_2bins.epsi, width=0.41\textwidth, angle=90}\hspace*{8mm}
\psfig{file=rabc_eta_2bins.epsi, width=0.41\textwidth, angle=90}\\[6mm]
\psfig{file=rabc_2bins1.epsi, width=0.41\textwidth, angle=90}\hspace*{8mm}
\psfig{file=rabc_eta_2bins1.epsi, width=0.41\textwidth, angle=90}
   \caption{{\it Ratio $r_{dA}$ (defined in Eq.~(\ref{rdef})) 
of double-parton and leading-twist contributions
in $dA\to\pi^0\pi^0X$.
The plots in the upper row are for $1.5< \eta_2 < 2$, the ones
in the lower row for $2.5< \eta_2 < 4$. For all plots we have 
chosen $1.5~\mathrm{GeV}<p_{T,2}<p_{T,1}$. 
The inserts show the ratios $r_a/r_c$ and $r_c/r_b$.}}
    \label{Rabc}
 \end{figure}

Figure~\ref{Rabc} shows the sum
\begin{equation}
r_{dA} \equiv r_a + r_b + r_c 
\label{rdef}
\end{equation}
at impact parameter $b=0$. It gives the ratio of the
full double-scattering contribution to the leading-twist one~\footnote{
Note that our $r_{dA}$ is not to be confused with the usual nuclear 
modification factor $R_{dA}$ mentioned in Sec.~\ref{3.1}.}.
One can see that double-parton interactions
in $dA$ scattering appear to lead to very significant enhancements of the 
cross section over the leading-twist one, much stronger than in $pp$
scattering. The inserts in the figure show the corresponding 
ratios $r_a/r_c$ and $r_c/r_b$ which show the trend discussed above.

\subsection{Impact on interpretation of pion azimuthal correlations\label{3.3}}

We expect our findings in Fig.~\ref{Rabc} to be also relevant for
the interpretation of the azimuthal distributions of the pions
mentioned earlier. Such distributions have recently been investigated
by the STAR and Phenix experiments at RHIC~\cite{rhicfwd2,starqm09,phenixqm09}.
What is measured is the distribution in the difference $\Delta\varphi$
of the azimuthal angles of the two pions. The distributions are
normalized relative to the total number of trigger events, 
that is, given a high-$p_{T,1}$ pion with rapidity $\eta_1$ 
that passes the selection cuts,
the $\Delta\varphi$ distribution gives the probability for 
finding a second pion in a given azimuthal bin. The (still 
preliminary) data show peaks corresponding to near-side 
($\Delta\varphi\sim 0$) and away-side ($\Delta\varphi\sim \pi$) 
correlations, on top of a broad ``pedestal'' that extends over 
all $\Delta\varphi$. The pedestal is significantly higher in 
$dA$ than in $pp$ scattering. Also it is found that in central $dA$ the 
away-side peak is strongly depleted when both pions are produced at 
forward rapidities, $\eta_i\sim 3$~\cite{rhicfwd2,starqm09,phenixqm09}. 

In view of the relatively early stage the data are in, 
our discussion will be overall be more qualitative here. 
Also, the theoretical framework is not sufficiently developed
for a full quantitative study. The leading-twist calculations 
we have done in the previous section were entirely in the framework 
of collinear factorization. Here the ingredients for a full calculation 
are essentially available, including next-to-leading order corrections 
(even though for simplicity we did not use these). In the case of the 
correlation function in $\Delta \varphi$, however, the calculation is 
much more involved. Away from $\Delta \varphi=\pi$, the leading-twist 
part will be dominated by $2\to 3$ processes, which are available. 
However, near $\Delta \varphi=\pi$ -- the most interesting region --
any finite order of perturbation theory will fail because of the presence 
of large Sudakov double-logarithms. A resummation of these logarithms to all
orders in perturbation theory is required here, which unfortunately so far has 
not been worked out. To perform this resummation 
is of course well outside the scope of this paper. 
In addition, also non-perturbative contributions will be present very close 
to $\Delta \varphi=\pi$. It seems to us that none of the theoretical studies 
of the correlation function addresses these contributions at an appropriate 
level. We could follow a standard procedure and attempt to model 
perturbative and non-perturbative contributions to the $\Delta \varphi$ 
correlation function using Gaussian smearing in parton transverse momenta; 
however, we refrain from such a rather ad-hoc approach and stick to 
a more qualitative discussion that captures the main physics.

Our first observation is that, as discussed in Sec.~\ref{sect21}, 
for the leading-twist mechanism 
the two pions will predominantly be produced back-to-back in azimuthal angle,
that is, around $\Delta\varphi=\pi$. Pure $2\to 2$ scattering produces
the pions at $\Delta\varphi=\pi$; the region away from the 
backward peak, around say $\Delta\varphi \sim \pi/2$, can only be 
filled by $2\to 3,4,\ldots$ scattering, which are of higher
order in the strong coupling $\alpha_s$.
These features are in contrast to the double-scattering mechanism,
for which the two pions are produced essentially uncorrelated in
$\Delta\varphi$ and which hence is expected to uniformly fill the 
$\Delta\varphi$ distribution. Since we found in the previous subsections
that double-scattering is prevalent at forward angles in 
$pp$ and in particular $dA$ scattering, we conclude that the 
{\it numerator} of the pedestal around $\Delta\varphi \sim \pi/2$ 
should be almost entirely due to the double-scattering mechanism. 
The {\it denominator}, on the other hand, is the single-inclusive
``trigger'' cross section $d^2\tilde{\sigma}/dp_{T,1}d\eta_1$, 
for which we can safely assume that
double-scattering contributions play a less important role. 
In any case, its value is known from past STAR measurements in 
$pp$ and $dA$ scattering~\cite{rhicfwd2}. Therefore,
the height of the pedestal for $dA$ is generically given as
\begin{equation}
{\mathrm{Ped}}_{\,dA} \approx 
\frac{d^4\sigma^{dA}_{\mathrm{double}}}{dp_{T,1}d\eta_1dp_{T,2}d\eta_2}
\;\left/\; \frac{d^2\tilde{\sigma}^{dA}}{dp_{T,1}d\eta_1},\right.
\label{pedestal}
\end{equation}
and similarly for $pp$ scattering. 
If our considerations are correct, we can estimate
the relative heights of the pedestals in $pp$ and $dA$ scattering:
\begin{equation}
\frac{{\mathrm{Ped}}_{\,dA}}{{\mathrm{Ped}}_{\,pp}} \approx
\frac{d^4\sigma^{dA}_{\mathrm{double}}}
{d^4\sigma^{pp}_{\mathrm{double}}}\times
\frac{d^2\tilde{\sigma}^{pp}}{d^2\tilde{\sigma}^{dA}}
=\frac{d^4\sigma^{dA}_{\mathrm{LT}}}{d^4\sigma^{pp}_{\mathrm{LT}}}
\times\frac{r_{dA}}{r_{pp}}
\times
\frac{d^2\tilde{\sigma}^{pp}}{d^2\tilde{\sigma}^{dA}},
\label{ped1}
\end{equation}
where in the second step we have used Eqs.~(\ref{eq7}) and (\ref{rdef}) 
to introduce the ratio of leading-twist and double-scattering
contributions. The last factor, $d^2\tilde{\sigma}^{pp}/
d^2\tilde{\sigma}^{dA}$, corresponds to the inverse of the nuclear 
modification
factor $R_{dA}$ that we mentioned earlier, at trigger transverse momentum
$p_{T,1}$ and rapidity $\eta_1$. We assume that the first 
factor, $d^4\sigma^{dA}_{\mathrm{LT}}/d^4\sigma^{pp}_{\mathrm{LT}}$, 
roughly shows the square of this suppression and hence is of 
order $R_{dA}^2$. We shall give a better argument for this below.
Then, one factor of $R_{dA}$ cancels in the ratio of the pedestals, and
we obtain
\begin{equation}
\frac{{\mathrm{Ped}}_{\,dA}}{{\mathrm{Ped}}_{\,pp}} 
\approx R_{dA} \times \frac{r_{dA}}{r_{pp}}\approx
3 \;.
\label{rest}
\end{equation}
Here we have used a typical value of $r_{dA}/r_{pp}\sim 10$ 
from our previous figures~\ref{MIfig} and~\ref{abc}, 
and~\cite{rhicfwd1,rhicfwd2,rhicfwd3}
 $R_{dA}\sim 1/3$. Obviously, the value we 
find in~(\ref{rest}) can only be a rough estimate; however, we
are encouraged by the fact that it is well consistent even 
quantitatively with the experimental observation of a significant 
enhancement of the pedestal in central $dA$ scattering~\cite{starqm09}.
Thus we conclude that the RHIC experiments may well have discovered the 
first example of multiparton interactions in many-nucleon systems, with 
all previous observations having been restricted to $pp$ or 
$p\bar p$  collisions.
Data with a finer binning in $\eta_1,\eta_2$ would allow a more detailed 
check of our expectations.

We can now go one step further and consider the away-side
peak around $\Delta\varphi \sim \pi$. In the peak region,
the structure of the two-pion correlation in $dA$ scattering is
\begin{equation}
{\mathrm{Peak}}_{\,dA} \approx \left(\frac{
d^4\sigma^{dA}_{\mathrm{LT}}}{dp_{T,1}d\eta_1dp_{T,2}d\eta_2}+
\frac{d^4\sigma^{dA}_{\mathrm{double}}}{dp_{T,1}d\eta_1dp_{T,2}d\eta_2}\right)
\;\left/\; \frac{d^2\tilde{\sigma}^{dA}}{dp_{T,1}d\eta_1}.\right. 
\end{equation}
As indicated, we here expect to have a contribution also from the
leading-twist term. We now subtract the pedestal term given
in Eq.~(\ref{pedestal}) and find
\begin{equation}
{\mathrm{Peak}}_{\,dA} -{\mathrm{Ped}}_{\,dA} \approx \frac{
d^4\sigma^{dA}_{\mathrm{LT}}}{dp_{T,1}d\eta_1dp_{T,2}d\eta_2}
\;\left/\; \frac{d^2\tilde{\sigma}^{dA}}{dp_{T,1}d\eta_1}.\right. 
\end{equation}
Taking again the ratio to the corresponding quantity in $pp$
scattering we obtain
\begin{equation}
\frac{{\mathrm{Peak}}_{\,dA} -{\mathrm{Ped}}_{\,dA}}{{\mathrm{Peak}}_{\,pp} 
-{\mathrm{Ped}}_{\,pp}}\approx 
\frac{d^4\sigma^{dA}_{\mathrm{LT}}}{d^4\sigma^{pp}_{\mathrm{LT}}}\times
\frac{d^2\tilde{\sigma}^{pp}}{d^2\tilde{\sigma}^{dA}}.
\label{peak}
\end{equation}
Compared to the pedestal ratio given in~(\ref{ped1}), this value
does not contain the factor $r_{dA}/r_{pp}\sim 10$. We hence conclude
that the {\it height of the peak above the pedestal} is about 
$R_{dA}\sim 1/3$ times smaller in $dA$ scattering than in $pp$. Compared 
to the relative heights of the pedestals in $dA$ and $pp$, this is a
reduction of even a factor 10. Again, both these findings are consistent 
with the observations in the data. Evidently, within our more
qualitative discussion one cannot rule out that there could also 
be a contribution due to the $2\to 1$ broadening mechanism discussed 
in \cite{Tuchin:2009nf,Albacete:2010pg} to the pion azimuthal correlation,
which in these models constitutes only a small fraction  ($\lesssim 1/6$) 
of the pedestal events. In view of the much larger double-parton mechanism 
contribution that we find for the same  $\Delta \varphi$, it is
not clear at the moment how to check experimentally to what degree 
such a $2\to 1$ broadening contribution is present.

The step that remains is to investigate the ratios 
$d^2\tilde{\sigma}^{pp}/d^2\tilde{\sigma}^{dA}$ and 
$d^4\sigma^{dA}_{\mathrm{LT}}/d^4\sigma^{pp}_{\mathrm{LT}}$ in~(\ref{ped1})
and~(\ref{peak}). The former is given by the nuclear modification
factor $R_{dA}$ which, as we mentioned, has been found experimentally
at RHIC to show a significant suppression of the $dA$ cross section
relative to the $pp$ one at forward rapidities. The mechanism behind 
this suppression is not conclusively understood so far. As discussed
in~\cite{GSV}, it could arise, at leading-twist level, from a combination 
of two effects. The first is the leading-twist shadowing 
phenomenon~\cite{Guzey:2009jr} whose 
impact on $R_{dA}$ we computed in~\cite{GSV}. Using the nuclear parton 
distribution functions (nPDFs) of~\cite{FGS} we found relatively small 
shadowing effects, because the relevant gluon momentum fractions 
in the ``target'' are on average not very small for single-inclusive
hadron production, $\left< x_g\right>\sim 0.02$, even at forward rapidities.
For such $x_g$, gluon shadowing is predicted in~\cite{FGS} to be relatively 
moderate. That said, little is known experimentally about gluon shadowing,
and the recent set~\cite{Salg} of nPDFs proposes a stronger shadowing 
effect. For our present study we will stick to the use of the nPDFs
of~\cite{FGS}. 

The second effect is energy loss of partons. It was shown
in~\cite{Martin} that partons propagating through the ``target'' 
nucleus in kinematics close to the black disk regime suffer ``fractional''
energy losses. The interactions near the black disk regime select 
configurations in which the parton has split into two or more 
partons. In \cite{GSV} we pointed out that even a relatively small 
energy loss of order $5$ to $10\%$, which is consistent with 
the estimated magnitude of this effect~\cite{FS07}, can explain 
the observed patterns of suppression in forward $dA$-scattering
at RHIC. Energy loss effects are also typically embodied in an 
effective way in ``nuclear-modified'' fragmentation functions; see 
for example~\cite{review,ssz,wang,armesto,Kopeliovich:2008uy}, 
which are fitted 
or compared to RHIC $dA$ and $AA$ data. These may hence also serve
as useful tools for investigating suppression effects in a leading-twist
calculation of single-inclusive or double-inclusive particle production 
at RHIC. 

 \begin{figure}[t]
\hspace*{1.2cm}  
\psfig{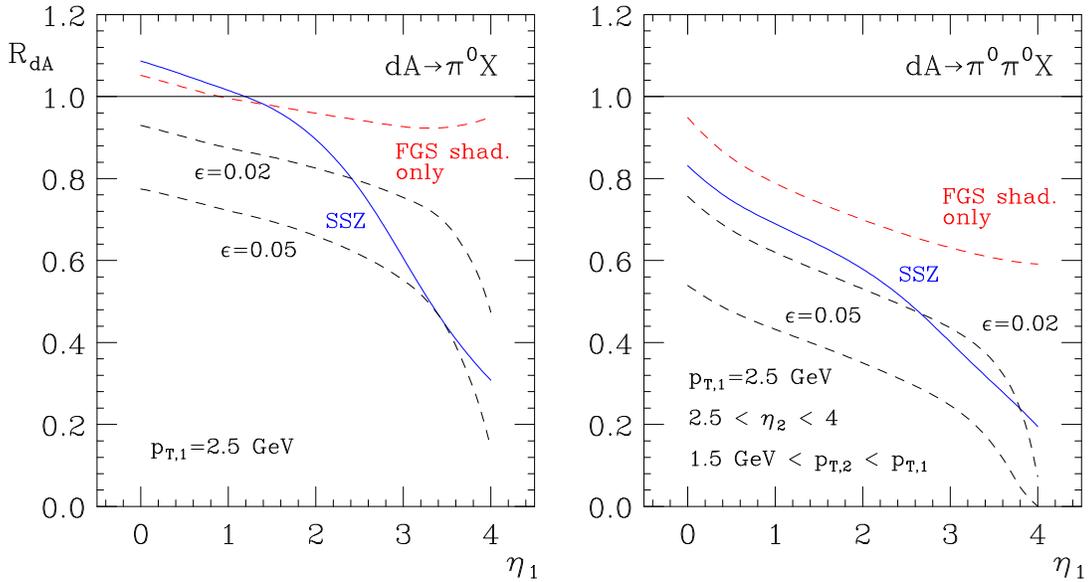}
   \caption{{\it Left: Nuclear modification factor $R_{dA}$ 
for single-inclusive leading-twist pion production as a function of 
rapidity $\eta_1$ at $p_{T,1}=2.5$~GeV. The upper dashed line shows 
the effect of leading-twist shadowing for the Frankfurt-Guzey-Strikman
(FGS) nuclear parton distributions~\cite{FGS}. The solid line 
includes shadowing and the ``medium-modified'' fragmentation functions
of Sassot-Stratmann-Zurita (SSZ)~\cite{ssz}. The lower dashed lines 
show the results for two simple energy-loss models, see text. 
Right: Same for double-inclusive pion production.  \label{figeloss}}}
 \end{figure}

The left part of Fig.~\ref{figeloss} makes our observations
more quantitative. We show results for $R_{dA}$ (for $A=Au$), 
computed from the leading-twist single-inclusive cross section
$d\tilde{\sigma}^{dA}_{\mathrm{LT}}/p_{T,1}d\eta_1$ in Eq.~(\ref{eq3})
and normalized to the 
corresponding $pp$ cross section. The upper dashed line shows the effect
of including leading-twist shadowing of~\cite{FGS} which, as discussed above,
is quite small. The solid line shows the result when using the same
shadowing and in addition the set of nuclear-modified fragmentation 
functions of Ref.~\cite{ssz}\footnote{We note that only {\it NLO} sets
of nuclear-modified fragmentation functions have been presented in the 
paper~\cite{ssz}. In order to avoid any mismatch with the DSS
set~\cite{DSS} that we use for the ordinary fragmentation functions, 
the solid curves in Fig.~\ref{figeloss} have been computed by
also using the {\it NLO} set of~\cite{DSS} for the calculation
of the $pp$ cross section in the denominator of $R_{dA}$.}. 
As one can see, $R_{dA}$ is suppressed,
except for the mid-rapidity region, where anti-shadowing effects
are relevant. The suppression grows with $\eta_1$ and is 
of order $1/3- 1/2$ at forward rapidities
of the pion, in line with the experimental observations.
This is expected since the fragmentation functions of ~\cite{ssz}
have been fitted to the RHIC forward single-pion production data. 

Interestingly, a simple model of energy loss generically yields 
results of the same size, as shown by the
two lower dashes lines. Here we have, in the spirit of the earlier
discussion of ``fractional'' energy loss, simply rescaled the 
momentum fraction $x_b$ of the parton in the gold nucleus by
$x_b\to x_b(1+\epsilon)$, and similarly for the fragmenting parton. 
At forward rapidities, the results for $\epsilon=0.02$ and $\epsilon=0.05$ 
roughly span the one obtained for the nuclear-modified fragmentation 
functions. At mid-rapidity, they are lower and fail to reproduce
the anti-shadowing effects seen in the data. This is not a surprise, 
however, since our simple energy loss estimate is only expected
to work at larger rapidities where the produced parton has to
traverse the largest amount of strongly-interacting matter
and where one is closer to the black disk regime. 
What is surprising is that even rather small values of $\epsilon$ 
generate relatively large suppression effects. This suggests that,
regardless of its precise mechanism, energy loss will always 
be expected to play a significant role in $dA$ scattering.
Note that in our picture energy losses are fractional only in  
the proximity of the black disk regime. Consequently, for fixed 
transverse momentum and increasing rapidity we expect 
$\epsilon $ to increase. In our rough estimates we have neglected 
this effect which obviously will work to amplify further the 
suppression effect. At the same time the energy losses are expected 
to be energy independent 
far away from the black disk regime~\cite{Dokshitzer}, 
which explains the absence of suppression of the forward-central 
correlations. We point out that the resummation of nuclear-enhanced
power corrections to the leading-twist cross section also results 
effectively in a shift of the momentum fraction of the initial
``projectile'' quark~\cite{qiuvitev}, whose size depends on kinematics. 
This approach was shown to be quantitatively consistent with the 
forward suppression of $R_{dA}$. 

Strikingly, the effects we find for single pions are amplified for
double-inclusive scattering. The corresponding results are shown
in the right part of Fig.~\ref{figeloss}. One reason for the 
additional suppression in this case is that significantly 
smaller momentum fractions are probed in the ``target'', down 
to $x_g\sim 10^{-3}$~\cite{GSV}, where gluon shadowing is stronger.
This effect is seen from the upper dashed line in Fig.~\ref{figeloss}. 
Furthermore, since two fragmentation functions are present for
double-inclusive pion production, the energy loss effect is much 
more prominent, as shown by the solid and lower dashed lines. 
Indeed, as we anticipated, the overall suppression of 
the $dA$ double-inclusive leading-twist cross section is  
roughly given by the square of that for the single-inclusive one, 
that is, by $R_{dA}^2$. This feature is likely 
generic for any kind of shadowing and energy loss mechanisms 
present in $dA$ scattering. In this sense, a strong depletion
of the backward peak in the pion azimuthal correlation in 
forward $dA$ scattering (which is of the same  magnitude as the  
experimentally observed depletion) is very natural, given the
previously found milder suppression of single-inclusive pion
production. It is of interest that the observed suppression of
the double-inclusive cross section at central impact parameters  
as compared to the impulse approximation is close 
to its lower bound corresponding to the probability that a quark 
passing through the nucleus encounters only one nucleon at its 
impact parameter, which for the case of scattering off gold at 
$b\sim 0$ is about $1/20 - 1/10$~\cite{Alvioli:2009ab}.

We note that for $pA$ scattering the dominant mechanism (b) of 
subsection~\ref{3.2} would be absent, so that the double-scattering 
contributions would remain a bit closer in size to what we found in 
the $pp$ case. The pion azimuthal correlation should then have 
a less pronounced pedestal, but a similarly suppressed backward peak.

We finally briefly address the ``near-side'' correlation of two
hadrons produced at $\Delta\varphi \sim 0$, both with large
rapidity. Experimentally, this correlation shows a strong peak
that (unlike the backward peak at $\Delta\varphi \sim \pi$) 
is not suppressed in $dA$ scattering. One may wonder 
what can be said in the context of our present calculation 
about the region $\Delta\varphi \sim 0$. It is clear that the 
near-side correlation receives contributions from a rather
different physics mechanism: the two hadrons produced at similar 
azimuthal angle and rapidity may originate from just a {\it single}
fragmentation process, through leading-twist double fragmentation 
of a high-$p_T$ final-state parton, described by
di-hadron fragmentation functions~\cite{dihadff}. Unfortunately,
little is known at present about the latter. 
Kinematically, such a contribution
is rather similar to a single-inclusive cross section. It is
therefore not expected to show the suppression $\sim R_{dA}^2$
that we found above for the leading-twist double-inclusive piece,
but rather a suppression of order $R_{dA}$. As a result, one
would naturally expect the near-side peak in $dA$ to show little 
suppression. To be more specific, we denote the contribution to
two-hadron production arising from double-fragmentation by
$d^4 \sigma_{\mathrm{df,LT}}^{dA}$. It adds to the (roughly 
$\Delta\varphi$-independent) double-scattering pedestal 
contribution. Subtracting the pedestal as before, we find
the following structure of the two-pion correlation in $dA$ scattering 
in the near-side peak region:
\begin{equation}
\left({\mathrm{Peak}}_{\,dA}-
{\mathrm{Ped}}_{\,dA}\right)_{\mathrm{near-side}}\; \approx \frac{
d^4\sigma^{dA}_{\mathrm{df,LT}}}{dp_{T,1}d\eta_1dp_{T,2}d\eta_2}
\;\left/\; \frac{d^2\tilde{\sigma}^{dA}}{dp_{T,1}d\eta_1}\right. \;,
\end{equation}
where as before $d^2\tilde{\sigma}^{dA}$ is the ordinary 
single-inclusive trigger 
piece. Taking again the ratio to the corresponding quantity in $pp$
scattering we obtain
\begin{equation}
\left(\frac{{\mathrm{Peak}}_{\,dA} -{\mathrm{Ped}}_{\,dA}}{{\mathrm{Peak}}_{\,pp} 
-{\mathrm{Ped}}_{\,pp}}\right)_{\mathrm{near-side}} \approx 
\frac{d^4\sigma^{dA}_{\mathrm{df,LT}}}{d^4\sigma^{pp}_{\mathrm{df,LT}}}\times
\frac{d^2\tilde{\sigma}^{pp}}{d^2\tilde{\sigma}^{dA}}\;,
\label{peak1}
\end{equation}
to be compared to Eq.~(\ref{peak}) for the away-side correlation at 
$\Delta\varphi \sim \pi$. {\it If} the double-fragmentation contribution
indeed behaves similar to a single-inclusive cross section-- which we
consider to be quite natural given its origin from fragmentation of
a single parton-- the two factors on the right-hand-side of Eq.~(\ref{peak1})
would be $R_{dA}$ and $1/R_{dA}$, respectively, and hence cancel. The whole
ratio would then be unity, which is indeed what the data~\cite{starqm09} show.
Even though this discussion is again rather qualitative, it offers
a natural and straightforward explanation of the non-suppression of 
the near-side peak in $dA$ scattering.

\section{Conclusions \label{sect4}}

We have investigated the role of double-scattering contributions
to double-inclusive pion production in $pp$ and $dA$ scattering
at RHIC. We have found that these become important at large
rapidities of the produced pions. This is in particular the case
for central $dA$ scattering, where the double-scattering contribution
can exceed the leading-twist one by large factors.  
Further detailed studies of double-inclusive pion production at RHIC
may provide a unique way for studying parton correlations in the nucleon. 
This is remarkable because traditionally only four-particle final states
were considered as possible probes of double-parton scattering. These
would typically not be viable at forward rapidities.

The double-scattering contributions appears to play a critical role 
in the interpretation
of the pion azimuthal correlations observed experimentally at RHIC. 
They primarily produce pions that are uncorrelated in azimuthal angle 
and hence are expected to strongly dominate the pedestals seen in the 
distributions. We have shown that the relative heights of the pedestals
in $pp$ and $dA$ scattering can be qualitatively understood in this way. 
We have furthermore shown that once the pedestal is subtracted,
the remaining backward correlation peak in $dA$ scattering is 
strongly affected by shadowing and energy loss effects. These are
found to be much stronger for double-inclusive scattering compared to 
single-inclusive, giving rise to a depletion of the backward peak 
in $dA$, consistent with the observations at RHIC.
Overall, in the light of our results, the patterns observed in the
pion azimuthal correlations at RHIC find a natural qualitative
explanation.

\section*{Acknowledgments}
We are grateful to  L.~Frankfurt, A.~Gordon, S.~Heppelmann, 
B.~Jacak, and L.~McLerran for useful discussions and comments. 
M.S. would like to thank the Yukawa International Program for 
Quark Hadron Sciences for hospitality and stimulating atmosphere 
during a part of this study. M.S.'s research was supported 
by DOE grant No. DE-FG02-93ER40771. W.V.'s work has been supported 
by the U.S. Department of Energy (contract number DE-AC02-98CH10886).

\newpage


\end{document}